\documentclass[journal]{IEEEtran}
\usepackage{bbm}
\usepackage{caption}
\usepackage{subfigure}
\usepackage{graphicx}
\usepackage{latexsym}
\usepackage{diagbox}
\usepackage{changepage}
\usepackage[fleqn]{amsmath}
\usepackage{amsmath}
\usepackage{amsfonts}
\usepackage{indentfirst}
\usepackage{CJK}
\usepackage{indentfirst}% Some very useful LaTeX packages include:
\usepackage[varg]{txfonts}% (uncomment the ones you want to load)
\usepackage{stfloats}% *** MISC UTILITY PACKAGES ***
\usepackage{multirow}%
\usepackage{booktabs}%\usepackage{ifpdf}
\usepackage{color,soul}% Heiko Oberdiek's ifpdf.sty is very useful if you need conditional
\usepackage{epstopdf}% compilation based on whether the output is pdf or dvi.
\usepackage{float}% usage:
\usepackage{bm}
\usepackage{makecell}
\usepackage{array}
\usepackage{bm}
\usepackage{mathtools}
\usepackage{geometry}
\usepackage[linesnumbered,ruled,commentsnumbered,longend]{algorithm2e}
\usepackage[linesnumbered,ruled,commentsnumbered,longend]{algorithm2e}
\makeatletter

\newcommand{\Rmnum}[1]{\expandafter\@slowromancap\romannumeral #1@}

\makeatother
\makeatletter
\newcommand{\qed}{\nobreak \ifvmode \relax \else
	\ifdim\lastskip<1.5em \hskip-\lastskip
	\hskip1.5em plus0em minus0.5em \fi \nobreak
	\vrule height0.75em width0.5em depth0.25em\fi}

\SetAlgoLongEnd

\ifCLASSINFOpdf
\else
\fi

\begin{document}

\title{\Large{Joint Beamforming And Power Splitting Design For C-RAN With Multicast Fronthaul}}
\author{Wanming Hao, Gangcan Sun, Ming Zeng, Zhengyu Zhu, Bin Jiang, and Shouyi Yang
%	\thanks{This work was supported in part by the National Key Research and Development Project under Grant 2019YFB1803200, in part by the National Natural Science Foundation of China under Grant 61801434, 61801435, U1604159, in part by the Project funded by China Postdoctoral Science Foundation under Grant 2019M662528, 2018M642784, in part by the Scientific and Technological Key Project of Henan Province under Grant 192102310178.  (\textit{Corresponding author: Zhengyu Zhu, Gangcan Sun})}
	\thanks{W. Hao is with the School of Information Engineering, and the Henan Institute of Advanced Technology, Zhengzhou University, Zhengzhou 450001, China, and  also with the National Center for International Joint Research of Electronic Materials and Systems, Zhengzhou 450001, China.  (Email: iewmhao@zzu.edu.cn).}
	\thanks{G. Sun, Z. Zhu, and S. Yang are with the School of Information Engineering, Zhengzhou University, Zhengzhou 450001, China. (E-mail: \{iegcsun, iezyzhu, iesyyang\}@zzu.edu.cn)}
	%\thanks{Z. Chu is with the 5G Innovation Centre, Institute of Communication Systems, University of Surrey, Guildford GU2 7XH, U.K. (E-mail: zheng.chu@surrey.ac.uk}
%	\thanks{O. Muta is with Center for Japan-Egypt Cooperation in Science and Technology, Kyushu University, Fukuoka 819-0395, Japan.  (E-mails: muta@\{ieee.org, ait.kyushu-u.ac.jp\})}
	\thanks{M. Zeng is with the Faculty of Science and Engineering, Laval University, Quebec, G1V0A6, Canada, and  also with the Faculty of Engineering and Applied Science, Memorial University, St. Johns, NL A1B 3X9, Canada. (E-mail: mzeng@mun.ca).}
		\thanks{B. Jiang is with the College of Information and Communication, National University of Defense Technology, Wuhan 430010, China. (E-mail: jiangbin17@nudt.edu.cn)}}




\maketitle
\begin{abstract}
In this paper, we investigate the joint beamforming and power splitting design problem in a base station (BS) cluster-based cloud radio access network (C-RAN) with multicast fronthaul, where users are jointly served by BSs within each cluster. Meanwhile, each user can simultaneously obtain information and energy from received signals. On this basis, under predefined minimum harvested energy of each user and maximum transmit power of each BS and central processor, we formulate a sum rate maximization problem by jointly optimizing multicast fronthaul beamforming, cooperative access beamforming and power split ratios. Due to the difficulty in solving  the formulated problem, we first transform it into a convex one by successive convex approximate and semidefinite program (SDP) relaxation techniques, and then propose an effective iterative algorithm. Moreover, we design a randomization method that can always obtain the rank-one solution. Finally, numerical results are conducted to validate the effectiveness of our proposed algorithm.  
\end{abstract}

\begin{IEEEkeywords}
SWIPT, C-RAN, Multicast.
\end{IEEEkeywords}

% \fi
\IEEEpeerreviewmaketitle

\section{Introduction}
To satisfy the increasing demands of the data rate in future mobile networks, ultra-dense base station (BS) deployment is deemed as one of effective schemes~\cite{1Duan}. However, this leads to more serious interference among BSs. To handle it, cloud radio access network (C-RAN) structure is developed, where each BS is connected to the central processor (CP)~\cite{2Peng}. The CP jointly manages the interference by global resource allocation, effectively improving the spectral efficiency (SE) and relieving the BSs' burden (via moving the baseband processing to the baseband unit (BBU) pool)~\cite{3Park}. However, the C-RAN structure brings another challenge, i.e, the selection of the fronthaul links carrier. Traditionally, wired fronthaul link  is adopted due to its high stability and capacity. Nonetheless, its high deployment cost makes it unsuitable for the ultra-dense BS, and thus, the wireless  carrier becomes a suitable candidate~\cite{4Yu}. 
In addition, coordinated multiple-point (CoMP) transmission technology is also an effective approach to remove the adjacent-BS interference, which has been widely applied in C-RAN, e.g.,~\cite{4Yu}-\cite{6Hu}. Since the cooperative BSs need to jointly serve users, the CP should transmit each user's signal to these BSs, i.e., {\textit{point-to-multipoint}} transmission. To realize the above,  multicast technique will be a suitable scheme, which has been adopted in~\cite{4Yu} and~\cite{6Hu}. Therefore, to further enhance the performance of system, CoMP transmission-based C-RAN with multicast fronthaul represents a promising solution.

On the other hand, the simultaneous wireless information and power transfer (SWIPT) technique is developed for improving energy efficiency, where both information and energy are extracted from the same received RF signals~\cite{7Zhang}.  Although a lot of recent works investigate the SWIPT, e.g.,~\cite{9Shi}-\cite{11Zeng}, the study under C-RAN structure is limited. The authors in~\cite{11a} consider the weighted sum rate maximization for a multiuser multiple-input multiple-output SWIPT system, where the beamformers in both the downlink and uplink, and the time allocation are jointly optimized. The authors in~\cite{12Shi} consider a transmit power minimization problem in a full duplex C-RAN, and four approaches of jointly optimizing beamformers, uplink transmit power and receiver power ratios are proposed. The authors in~\cite{12a} study the  min max fronthaul load optimization problem for an energy harvesting powered C-RAN with QoS and harvested energy constraints, and propose an effective beamforming algorithm to solve it. In~\cite{13Peng}, the authors investigate SWIPT problem in an uplink C-RAN, and the minimum mean-square-error  is considered by optimizing precoders and detectors. \cite{14Zhao}~considers an energy-efficient uplink resource allocation problem by optimizing the sub-carrier and power allocation, and then, a quantum-behaved particle swarm-based low-computational suboptimal algorithm is proposed.  The max-min fair SWIPT is investigated in a green C-RAN with millimeter wave (mmWave) fronthaul~\cite{15Chen}. The minimum data rate is maximized via a two-step iterative bemaforming  algorithm.  Although~\cite{12Shi}-\cite{15Chen} involve the SWIPT in C-RAN, the cooperation among multiple BSs is not considered. Therefore, the multicast fronthaul and CoMP techniques are not investigated, including the multicast beamforming and cooperative beamforming design.

{Unlike the previous works, in this paper, we consider a downlink BS cluster-based SWIPT C-RAN with multicast fronthaul, where multiple BSs jointly serve users within one cluster. In general, the fronthaul link distance between the CP and the BS cluster is relatively large, and thus, adopting mmWave  is not appropriate due to its large path loss. Therefore, microwave (current cellular frequency, e.g., sub-6 GHz) is adopted owing to its small path loss. In contrast, the access link distance between the BS cluster and users is small, and thus, using mmWave is appropriate.} Specifically, the CP transmits data to each BS cluster via microwave multicast fronthaul links, and users simultaneously receive information and energy from the BS via mmWave access links. Based on this, we formulate a sum rate maximization problem by jointly optimizing multicast fronthaul  beamforming, cooperative access beamforming and power split ratios, under maximum transmit power constraint for each BS and the CP. For the formulated non-convex optimization problem, we first transform it into a semidefinite program (SDP)  optimization problem. Then, via successive convex approximate (SCA) and SDP relaxation techniques, the SDP problem  is relaxed  into a convex one, and an iterative algorithm is proposed. Finally, we propose a rank-one solution based on the randomization method.
%
%\textit{Notations}: We use the following notations throughout this paper:
%$(\cdot)^H$ denote Hermitian transpose, $\|\cdot\|$ is the Frobenius norm, {Tr($\cdot$)} denotes trace operation.
\section{System Model and Problem Formulation}
We consider a downlink C-RAN with one  CP and $L$ BS clusters.
%as shown in Fig.~\ref{figure1}%. 
The CP is equipped with $N$ antennas, while each BS is mounted with one antenna for transmitting mmWave signal and receiving microwave signal simultaneously. We assume  that there are $M$ BSs and $K$ users in each cluster,
%\footnote{{Here, $M$ cooperating BSs can serve at $M$ users each time. If there are more users in the cluster, they can be served by time division multiplex access. Here, $K (K\leq M)$ denotes the number of users being served by BSs.}}
and each user is jointly served by $M$ BSs. It is assumed that each user is equipped with power splitter hardware that can split the received signal into the information decoder (ID) and energy harvester (EH). In addition, the beamforming and power splitting design takes place at the CP.
%\begin{figure}[t]
%	\begin{center}
%		\includegraphics[width=4cm,height=3cm]{Systemfigure.pdf}
%		\caption{BS cluster-based C-RAN with multicast fronthual.}
%		\label{figure1}
%	\end{center}
%\end{figure}

%\subsection{Microwave Multicast Fronthaul Link} 
The received signal at BS $(l,m)$ can be expressed as
\begin{eqnarray}
	y_{lm}^{\rm{FH}}={\bf{h}}_{lm}{\bf{v}}_lx_l+\sum\nolimits_{j\neq l}^{L}{\bf{h}}_{lm}{\bf{v}}_jx_j+n_{lm},
\end{eqnarray}
where BS $(l,m)$ denotes the $m$th BS of the $l$th cluster, ${\bf{h}}_{lm}\in\mathbb{C}^{1\times N}$ represents the fronthaul link channel coefficient from the CP to BS $(l,m)$, ${\bf{v}}_l\in\mathbb{C}^{N\times 1}$ is the multicast beamforming for the $l$th cluster, and $x_l$ is the multicast signal with $\mathbb{E}\{|x_l|^2\}=1$. ${\bf{n}}_{lm}$ is an independent and identically distributed (i.i.d.) additive white Gaussian noise (AWGN), where each entry follows~$\mathcal{CN}(0,\delta^2)$.

To this end, the achievable fronthaul rate can be written as
\begin{eqnarray}
	R_{lm}^{\rm{FH}}=B_{\rm{mc}}\log\left(1+\frac{|{\bf{h}}_{lm}{\bf{v}}_l|^2}{\sum_{j\neq l}^{L}|{\bf{h}}_{lm}{\bf{v}}_j|^2+B_{\rm{mc}}\delta^2}\right),
\end{eqnarray}
where $B_{\rm{mc}}$ denotes the microwave bandwidth. The multicast rate of the $l$th cluster is decided by the BS with the worst channel condition, and thus the fronthaul multicast rate provided by the CP for the $l$th cluster is given
~\cite{12a}
\begin{eqnarray}
	R_{l}^{\rm{FH}}=\underset{m\in{\mathcal{M}}}{\min}\;\;\;\left\{R_{lm}^{\rm{FH}}\right\}, l\in\mathcal{L},
\end{eqnarray}
where ${\mathcal{M}}=\{1,\cdots,M\}$ and $\mathcal{L}=\{1,\cdots,L\}$ denote the BS and cluster sets, respectively.

%\subsection{Millimeter Wave Access Link} 
The received signal of User ($l,k$) can be expressed as
\begin{eqnarray}\label{signal}
	y_{lk}^{\rm{AC}}={\bf{g}}_{llk}{\bf{w}}_{lk}x_{lk}+\sum_{i\neq k}^{K}{\bf{g}}_{llk}{\bf{w}}_{li}x_{li}+\sum_{j\neq l}^{L}\sum_{i=1}^{K}{\bf{g}}_{jlk}{\bf{w}}_{ji}x_{ji}+n_{lk},
\end{eqnarray}
where User ($l,k$) denotes the $k$th user of the $l$th cluster, ${\bf{g}}_{jlk}=[g_{jlk}^1,\cdots,g_{jlk}^M]$ represents the downlink channel coefficient from $M$ BSs of the $j$th cluster  to User ($l,k$), $g_{jlk}^m$ denotes the downlink channel coefficient from BS ($j,m$) to User ($l,k$). In addition, ${\bf{w}}_{lk}\in\mathbb{C}^{M\times 1}$ and $x_{lk}$, respectively, denote the cooperative beamforming and signal for User ($l,k$), and ${\bf{n}}_{lm}$ is an i.i.d. AWGN with  $\mathcal{CN}(0,\delta^2)$. In~(\ref{signal}), the first term stands for the desired signal, the second term is the intra-cluster interference, and the third term represents the inter-cluster interference.

The received signal at each user are divided into two parts, i.e., ID and EH. Let $\beta_{lk}$ denote the power splitting factor for User ($l,k$), the received signal for the ID can be expressed as
%\begin{eqnarray}
\setlength{\mathindent}{0cm}
\begin{eqnarray}
y_{lk}^{\rm{ID}}&=\sqrt{\beta_{lk}}y_{lk}^{\rm{AC}}+u_{l,k},
%\\
%&=\sqrt{\beta_{lk}}\left({\bf{g}}_{llk}{\bf{w}}_{lk}x_{lk}\!+\!\sum_{i\neq k}^{K}{\bf{g}}_{llk}{\bf{w}}_{li}x_{li}\!+\!\sum_{j\neq %l}^{L}\sum_{i=1}^{K}{\bf{g}}_{jlk}{\bf{w}}_{ji}x_{ji}\!+\!n_{lk}\right)\!+\!u_{l,k},\nonumber 
\end{eqnarray}	
%\end{eqnarray} 
where $u_{l,k}$ denotes the caused noise due to the power splitting and follows~$\mathcal{CN}(0,\delta^2_u)$~\cite{9Shi}. Accordingly, the achievable rate of  User ($l,k$) can be written as
\begin{eqnarray}
	R_{lk}^{\rm{AC}}=B_{\rm{mm}}\log\left(1+\gamma_{lk}\right),
\end{eqnarray}
where $B_{\rm{mm}}$ is downlink mmWave bandwidth, and 
\begin{eqnarray*}
\gamma_{lk}=\frac{|{\bf{g}}_{llk}{\bf{w}}_{lk}|^2}{\sum_{i\neq k}^{K}|{\bf{g}}_{llk}{\bf{w}}_{li}|^2+\sum_{j\neq l}^{L}\sum_{i=1}^{K}|{\bf{g}}_{jlk}{\bf{w}}_{ji}|^2+B_{\rm{mm}}\delta^2+\delta^2_u/\beta_{lk}}. 
\end{eqnarray*}
In addition, the received signal for the EH can be expressed~as
\begin{eqnarray}
	y_{lk}^{\rm{EH}}=\sqrt{1-\beta_{lk}}y_{lk}^{\rm{AC}},
\end{eqnarray}
and the harvested energy is
\begin{eqnarray}
   E_{lk}=\eta(1-\beta_{lk})\left(\sum\nolimits_{j=1}^{L}\sum\nolimits_{i=1}^{K}|{\bf{g}}_{jlk}{\bf{w}}_{ji}|^2+B_{\rm{mm}}\delta^2\right),
   \end{eqnarray}
where $\eta$ denotes the energy conversion efficiency at each user.
\subsection{Problem Formulation}
In this paper, our objective is to maximize the sum rate by jointly optimizing beamforming and power split ratios, which can be formulated as
\begin{subequations}\label{OptA}
	\begin{align}
	\;\;\;\;\;\;\;\;\underset{\{{\bf{v}}_l,{\bf{w}}_{l,k},\beta_{lk}\}}{\rm{max}}\;&\sum\nolimits_{l=1}^{L}\sum\nolimits_{k=1}^{K}R_{lk}^{\rm{AC}}\label{OptA0}\\
	{\rm{s.t.}}\;\;&E_{lk}\geq E_{lk}^{\min}, \forall l,k,\label{OptA1}\\
	&\sum\nolimits_{l=1}^{L}||{\bf{v}}_l||^2\leq P_{\rm{CP}}^{\max}, \label{OptA2}\\
	&\sum\nolimits_{k=1}^{K}|{\bf{w}}_{l,k}(m)|^2\leq P_{l,m}^{\max}, \forall l,m,\label{OptA3}\\
	&\sum\nolimits_{k=1}^{K}R_{lk}^{\rm{AC}}\leq R_{l}^{\rm{FH}}, \forall l,\label{OptA4}
		\end{align}
\end{subequations} 
where (\ref{OptA1}) denotes the minimum harvested energy for each user, (\ref{OptA2}) and (\ref{OptA3}) represent the maximum transmit power constraints for the CP and  BS ($l,m$), respectively, where ${\bf{w}}_{l,k}(m)$ is the $m$th element of ${\bf{w}}_{l,k}$, and (\ref{OptA4}) is the fronthaul capacity constraint. Due to the non-convex objective function~(\ref{OptA0}), constraints~(\ref{OptA1}) and~(\ref{OptA4}), (\ref{OptA}) is a non-convex optimization~problem.
\section{Proposed Solution}
{First, we define the semi-definite beamforming matrix ${\bf{V}}_l={\bf{v}}_l{\bf{v}}_l^H$ and  ${\bf{W}}_l={\bf{w}}_l{\bf{w}}_l^H$. Accordingly, the rank of ${\bf{V}}_l$ and ${\bf{W}}_l$ should be one, namely ${\rm{rank}}({\bf{V}}_l)=1$ and ${\rm{rank}}({\bf{W}}_{lk})=1$.} Via introducing  auxiliary variables $a_{lk}$, $b_{lk}$ and $c_{lk}$, (\ref{OptA}) can be recast as the following SDP optimization problem
\begin{subequations}\label{OptB}
	\begin{align}
	\;\;\;&\underset{\{{\bf{V}}_l,{\bf{W}}_{l,k},\beta_{lk},a_{lk}, b_{lk}, c_{lk}\}}{\rm{max}}\;\sum\nolimits_{l=1}^{L}\sum\nolimits_{K=1}^{K}B_{\rm{mm}}\log\left(1+a_{lk}\right)\label{OptB0}\\
	{\rm{s.t.}}\;\;&a_{lk}\leq \gamma_{lk}', \forall l, k,\label{OptB1}\\
	&\sum\nolimits_{j=1}^{L}\sum\nolimits_{i=1}^{K}{\rm{Tr}}({\bf{G}}_{jlk}{\bf{W}}_{ji})+B_{\rm{mm}}\delta^2\geq b_{lk},\forall l,k,\label{OptB2}\\
	&b_{lk}(1-\beta_{lk})\geq E_{lk}^{\min}/\eta, \forall l, k,\label{OptB3}\\
	&\sum\nolimits_{l=1}^{L}{\rm{Tr}}({\bf{V}}_{l})\leq P_{\rm{CP}}^{\max},\label{OptB4}\\
	&\sum\nolimits_{k=1}^{K}{\bf{W}}_{lk}(m,m)\leq P_{l,m}^{\max}, \forall l,m,\label{OptB5}\\
	&\sum\nolimits_{k=1}^{K}B_{\rm{mm}}\log\left(1+\gamma_{lk}'\right)\leq c_{l}, \forall l,\label{OptB6}\\
	&c_{l}\leq B_{\rm{mc}}\log\left(1+\frac{{\rm{Tr}}({\bf{H}}_{lm}{\bf{V}}_l)}{\sum_{j\neq l}^{L}{\rm{Tr}}({\bf{H}}_{lm}{\bf{V}}_j)+B_{\rm{mc}}\delta^2}\right), \forall l,m,\label{OptB7}\\
	&{\rm{rank}}({\bf{V}}_l)=1,{\rm{rank}}({\bf{W}}_{lk})=1, \forall l,k,\label{OptB8}\\
	&{\bf{V}}_l\succeq{\bf{0}},{\bf{W}}_{lk}\succeq{\bf{0}}, \forall l,k,\label{OptB9}
	\end{align}
\end{subequations} 
where  $\gamma_{lk}'=\frac{{\rm{Tr}}({\bf{G}}_{llk}{\bf{W}}_{lk})}{\sum_{i\neq k}^{K}{\rm{Tr}}({\bf{G}}_{llk}{\bf{W}}_{li})+\sum_{j\neq l}^{L}\sum_{i=1}^{K}{\rm{Tr}}({\bf{G}}_{jlk}{\bf{W}}_{ji})+B_{\rm{mm}}\delta^2+\delta^2_u/\beta_{lk}}$, ${\bf{G}}_{jlk}={\bf{g}}_{jlk}^H{\bf{g}}_{jlk}$, ${\bf{H}}_{lm}={\bf{h}}_{lm}^H{\bf{h}}_{lm}$.

One can observe that (\ref{OptB}) is still a non-convex optimization problem due to non-convex constraints (\ref{OptB1}), (\ref{OptB3}), (\ref{OptB6}), (\ref{OptB7}) and rank-one constraint~(\ref{OptB8}). Next, we will transform them into the convex ones by advanced approximated technologies. We first introduce auxiliary variables $\xi_{lk}$ and $\epsilon_{lk}$, and (\ref{OptB1}) can be split into the following constraints
\begin{subequations}
\begin{align}
	\;\;\;\;\;\;\;\;\;\;\;\;\;\;\;\;\;\;\;\;a_{lk}\xi_{lk}&\leq {\rm{Tr}}({\bf{G}}_{llk}{\bf{W}}_{lk}),\forall l, k,\label{B11}\\
\xi_{lk}&\geq \Gamma_{lk}+B_{\rm{mm}}\delta^2+\delta^2_u\epsilon_{lk},\forall l, k,\label{B12}\\
\epsilon_{lk}&\geq 1/\beta_{lk},\forall l, k,\label{B13}
\end{align}
\end{subequations}
where $\Gamma_{lk}=\sum_{i\neq k}^{K}{\rm{Tr}}({\bf{G}}_{llk}{\bf{W}}_{li})+\sum_{j\neq l}^{L}\sum_{i=1}^{K}{\rm{Tr}}({\bf{G}}_{jlk}{\bf{W}}_{ji})$. In addition, the upper bound of $a_{lk}\xi_{lk}$ can be expressed as
\begin{eqnarray}\label{upA}
	{a_{lk}^{[n]}}\xi_{lk}^2/{2\xi_{lk}^{[n]}}+{\xi_{lk}^{[n]}}a_{lk}^2/{2a_{lk}^{[n]}}\geq a_{lk}\xi_{lk},\forall l, k,
\end{eqnarray}
where $a_{lk}^{[n]}$ and $\xi_{lk}^{[n]}$, respectively, denote the values of $a_{lk}^{[n]}$ and $\xi_{lk}^{[n]}$ at the $n$th iteration, and thus
(\ref{B11}) can be formulated into the following convex constraint
\begin{eqnarray}\label{convex1}
{a_{lk}^{[n]}}\xi_{lk}^2/{2\xi_{lk}^{[n]}}+{\xi_{lk}^{[n]}}a_{lk}^2/{2a_{lk}^{[n]}}\leq {\rm{Tr}}({\bf{G}}_{llk}{\bf{W}}_{lk}),\forall l, k.
\end{eqnarray}

Next, we can transform (\ref{OptB3}) and (\ref{B13}) into the following linear matrix inequality (LMI) constraints
  \begin{eqnarray}\label{LMI}
  	\left[ \begin{array}{ccc}
  	b_{lk} & \sqrt{E_{lk}^{\min}/\eta} \\
  	\sqrt{E_{lk}^{\min}/\eta} & (1-\beta_{lk}) 
  	\end{array} 
  	\right ]\succeq {\bf{0}},\;\;\left[ \begin{array}{ccc}
  	\epsilon_{lk} & 1\\
  	1 & \beta_{lk} 
  	\end{array} 
  	\right ]\succeq {\bf{0}},\forall l, k.
  \end{eqnarray}
  
To handle~(\ref{OptB6}), we introduce auxiliary variables $d_{lk}$, $\upsilon_{lk}$, $\tau_{lk}$, and split it into the following constraints
\begin{subequations}
	\begin{align}
	\;\;\;\;\;\;\;\;\;\;\;\;\;\;\;\;\;\;\;\;c_{l}&\geq \sum\nolimits_{k=1}^{K}B_{\rm{mm}}\log\left(1+d_{lk}\right),\forall l,\label{B21}\\
  \upsilon_{lk}^2&\leq 	d_{lk}(\Gamma_{lk}+B_{\rm{mm}}\delta^2+\delta^2_u\tau_{lk}),\forall l, k,\label{B22}\\
	\tau_{lk}\beta_{lk}&\leq 1 ,\forall l, k,\label{B23}\\
     \upsilon_{lk}^2&\geq \phi_{lk},\forall l, k,\label{B24}\\
     \phi_{lk}&\geq {\rm{Tr}}({\bf{G}}_{llk}{\bf{W}}_{lk}),\forall l, k.\label{B25}
	\end{align}
\end{subequations}
{By first-order Taylor approximation technique, we have $\log(1+d_{lk})\approx \log(1+d_{lk}^{[n]})+(d_{lk}-d_{lk}^{[n]})/(1+d_{lk}^{[n]})$, where $d_{lk}^{[n]}$ denotes the value of $d_{lk}$ at the $n$th iteration. Then, (\ref{B21}) can be transformed into the following convex constraint:}
\begin{eqnarray}\label{B3}
	c_{l}\geq\sum\nolimits_{k=1}^{K}B_{\rm{mm}}\left(\log\left(1+d_{lk}^{[n]}\right)+\frac{d_{lk}-d_{lk}^{[n]}}{1+d_{lk}^{[n]}}\right),\forall l,
\end{eqnarray}
where $d_{lk}^{[n]}$ denote the value of $d_{lk}$ at the $n$th iteration. (\ref{B22}) can be reformulated the following LMI constraint
\begin{eqnarray}\label{B4}
	\left[ \begin{array}{ccc}
	d_{lk} & \upsilon_{lk} \\
	\upsilon_{lk} & \Gamma_{lk}+B_{\rm{mm}}\delta^2+\delta^2_u\tau_{lk}
	\end{array} 
	\right ]\succeq {\bf{0}},\forall l, k.
\end{eqnarray}

Similar to~(\ref{upA}) and (\ref{convex1}), (\ref{B23}) can be approximated as the following convex constraint 
\begin{eqnarray}\label{upB}
{\tau_{lk}^{[n]}}\beta_{lk}^2/{2\beta_{lk}^{[n]}}+{\beta_{lk}^{[n]}}\tau_{lk}^2/{2\tau_{lk}^{[n]}}\leq 1,\forall l, k.
\end{eqnarray}
where $\tau_{lk}^{[n]}$ and $\beta_{lk}^{[n]}$, respectively, denote the values of $\tau_{lk}^{[n]}$ and $\beta_{lk}^{[n]}$ at the $n$th iteration. In addition, according to the first-order Taylor approximation,  $\upsilon_{lk}^2$ can be expressed as
\begin{eqnarray}
	\upsilon_{lk}^2\geq 2\upsilon_{lk}^{[n]}\upsilon_{lk}-(\upsilon_{lk}^{[n]})^2,\forall l, k,
\end{eqnarray} 
and we can transform (\ref{B24}) as the following convex constraint
\begin{eqnarray}\label{convex2}
	 2\upsilon_{lk}^{[n]}\upsilon_{lk}-(\upsilon_{lk}^{[n]})^2\geq \phi_{lk},\forall l, k.
\end{eqnarray}

\begin{figure*}[t]
	\centering
	\subfigure[]{
		\label{Iteration} %% label for first subfigure
		\includegraphics[width=5.3cm,height=4.5cm]{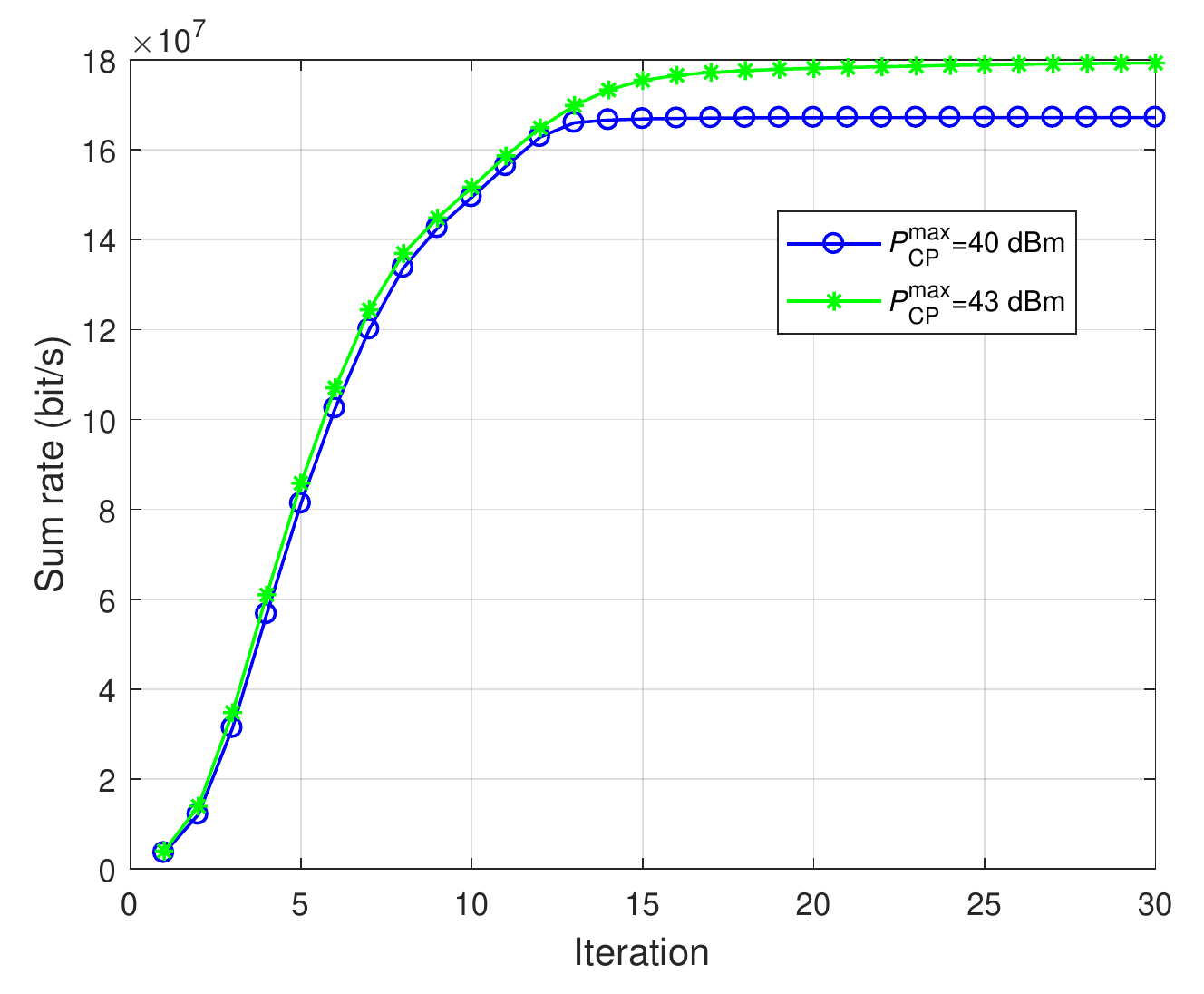}}
	%\hspace{1in}
	\subfigure[]{
		\label{CPPower} %% label for second subfigure
		\includegraphics[width=5.3cm,height=4.5cm]{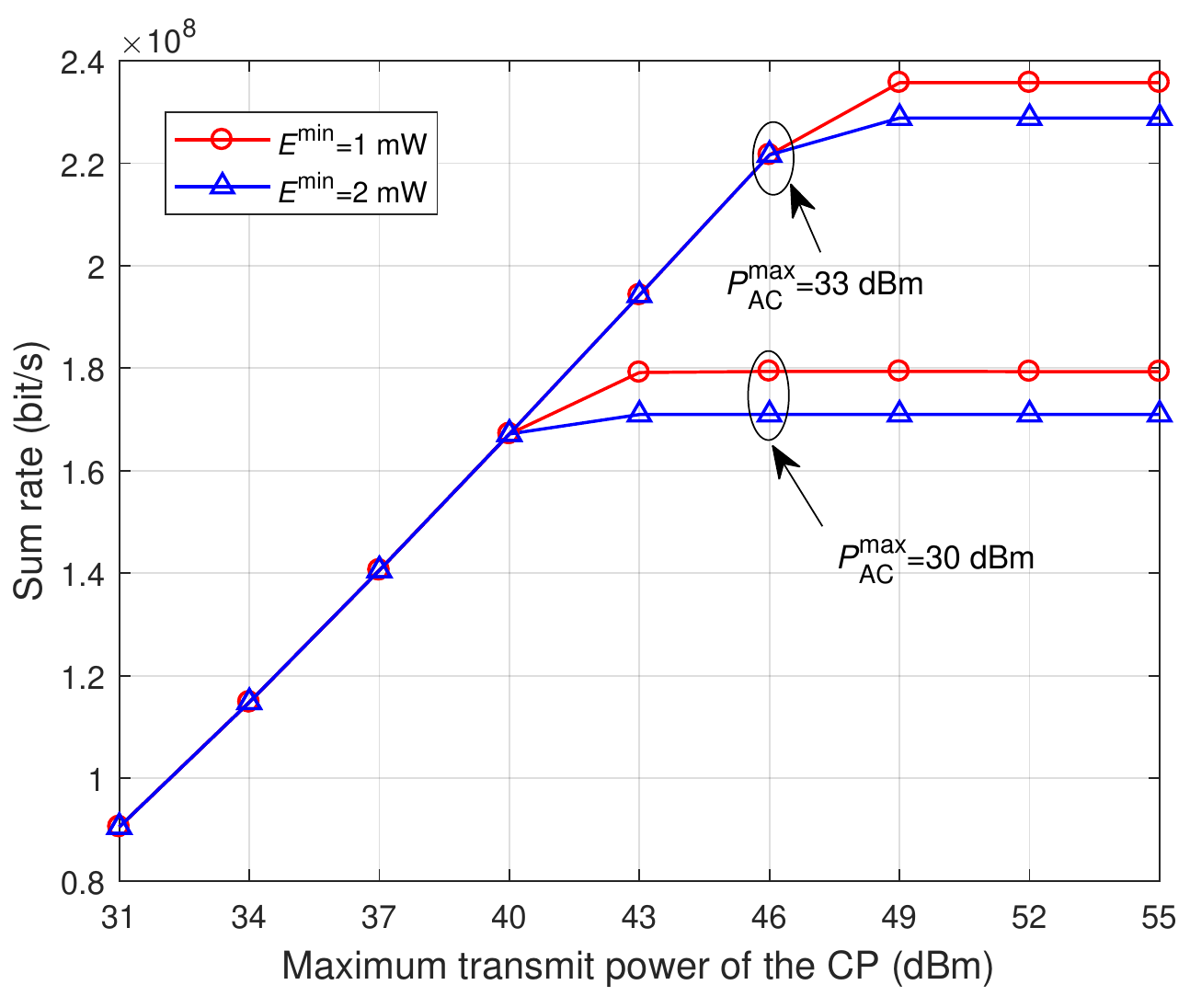}}
	\subfigure[]{
		\label{BSPower} %% label for second subfigure
		\includegraphics[width=5.3cm,height=4.5cm]{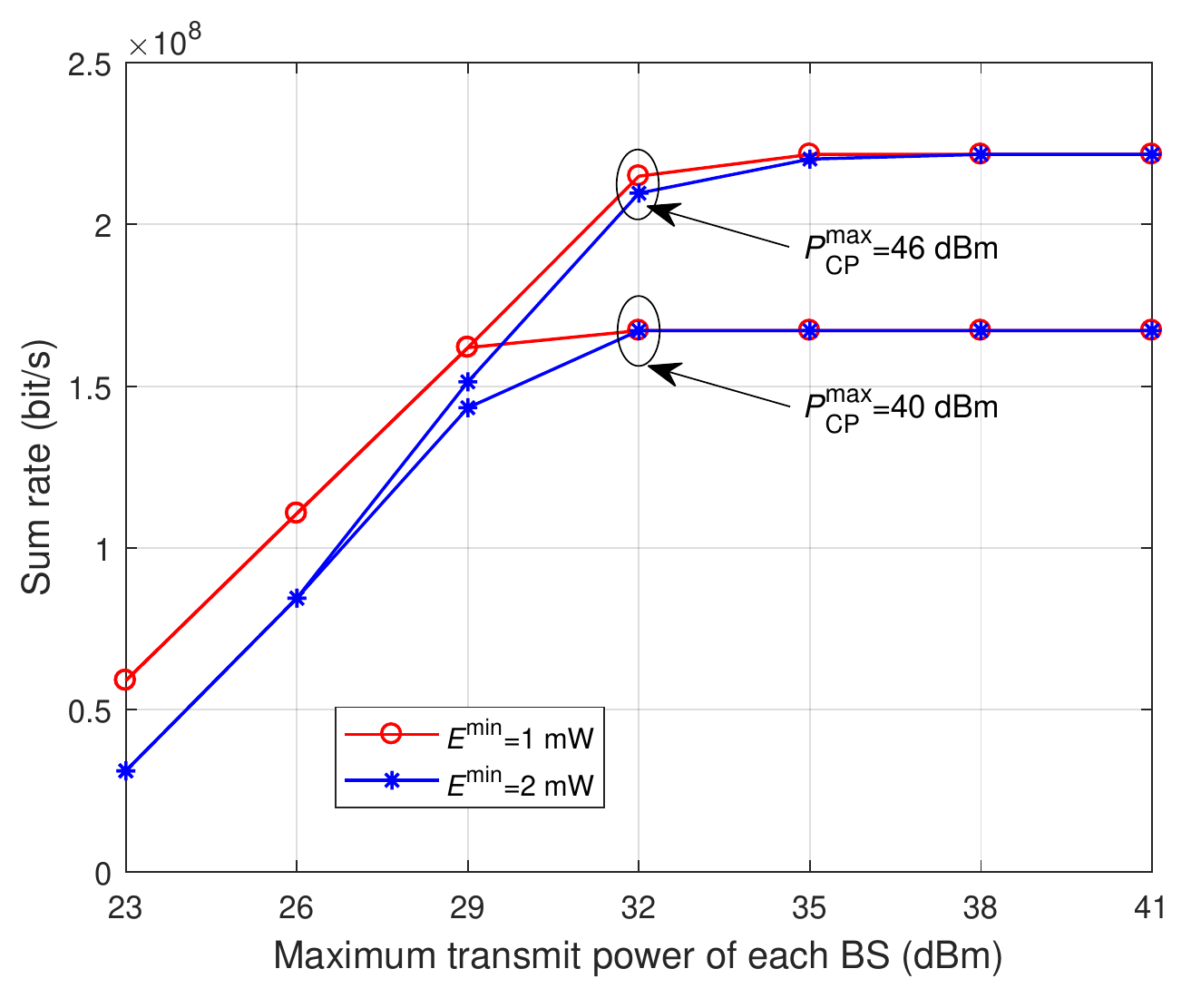}}
	\caption{Sum rate versus (a) iterations, (b) maximum transmit power of the CP, and (c) maximum transmit power of each BS.}
	%\label{fig:subfig} %% label for entire figure
\end{figure*}

Finally, we need to handle with (\ref{OptB7}). By introducing auxiliary variables $\lambda_{lm}$, $\omega_{lm}$, (\ref{OptB7}) can be split into as
\begin{subequations}
	\begin{align}
		\;\;\;\;\;\;\;\;\;\;\;\;\;\;\;\;\;\;\;\;\;\;c_l&\leq B_{\rm{mc}}\log(1+\lambda_{lm}),\forall l,m,\label{B41}\\
		\lambda_{lm}\omega_{lm}&\leq {\rm{Tr}}({\bf{H}}_{lm}{\bf{V}}_l),\forall l,m,\label{B42}\\
		\omega_{lm}&\geq \sum\nolimits_{j\neq l}^{L}{\rm{Tr}}({\bf{H}}_{lm}{\bf{V}}_j)+B_{\rm{mc}}\delta^2,\forall l,m.\label{B43}
\end{align}
\end{subequations}

One can observe that only (\ref{B42}) is non-convex constraint. Similar to the previous method, we directly transform (\ref{B42}) into the following convex constraint
\begin{eqnarray}\label{convex3}
	{\lambda_{lm}^{[n]}}\omega_{lm}^2/{2\omega_{lm}^{[n]}}+{\omega_{lm}^{[n]}}\lambda_{lm}^2/{2\lambda_{lm}^{[n]}}\leq {\rm{Tr}}({\bf{H}}_{lm}{\bf{V}}_l),\forall l,m.
\end{eqnarray}

Now, the only obstacle is the rank-one constraint~(\ref{OptB8}). By SDP relaxation, i.e., removing (\ref{OptB8}), we can obtain the following convex relaxed SDP optimization problem
\setlength{\mathindent}{0cm}
\begin{subequations}\label{OptC}
	\begin{align}
	&\underset{\{{\bf{V}}_l,{\bf{W}}_{l,k},\beta_{lk},a_{lk}, b_{lk}, c_{lk},\xi_{lk},\epsilon_{lk},d_{lk},\upsilon_{lk},\tau_{lk},\lambda_{lm},\omega_{lm}\}}{\rm{max}}\;\sum\nolimits_{l=1}^{L}\sum\nolimits_{K=1}^{K}B_{\rm{mm}}\log\left(1+a_{lk}\right)\label{OptC0}\\
	&{\rm{s.t.}}\;\;{\rm{(\ref{OptB2}),(\ref{OptB4}),(\ref{OptB5}),(\ref{OptB9}),(\ref{B12}), (\ref{convex1}),(\ref{LMI}), (\ref{B25}),(\ref{B3})}},\nonumber\\
			&\;\;\;\;\;\;\;{\rm{(\ref{B4}),(\ref{upB}),(\ref{convex2}),(\ref{B41}),(\ref{B43}),(\ref{convex3})}}.
	\end{align}
\end{subequations}

Problem~(\ref{OptC}) can be solved by standard convex optimization technique, e.g., interior-point method~\cite{17Convex}. On this basis, to obtain the solution of the problem~(\ref{OptB}), we need to iteratively solve~(\ref{OptC}). Specifically, starting from an initial feasible solution, we update $\{a_{lk}^{[n]}, \xi_{lk}^{[n]},d_{lk}^{[n]},\tau_{lk}^{[n]},\beta_{lk}^{[n]},\upsilon_{lk}^{[n]},\lambda_{lm}^{[n]},\omega_{lm}^{[n]}\}$ iteratively by solving (\ref{OptC}) using the obtained results from the previous iteration. The above procedure is carried out until convergence. We summarize the above iterative scheme in Algorithm~1. {To evaluate the characteristic of  rank-one for the obtained solutions, we perform 1000 times random trials, and the number of rank-one solutions is 996 (99.6\%), which shows the efficiency of the proposed Algorithm 1. Meanwhile, when the obtained solution does not satisfy the rank-one characteristic, we propose a randomization method to obtain the rank-one solution, and the detailed procedure can be found in Appendix.}

{Next, we discuss the effectiveness of the proposed algorithm. To obtain the solutions of the original non-convex problem (\ref{OptA}), we need to iteratively solve the convex problem (\ref{OptC}). The optimal solutions of (\ref{OptC}) can be obtained at each iteration, since it is a convex optimization problem. Moreover, iteratively solving (\ref{OptC}) will increase or at least maintain the value of the objective function in (\ref{OptC})~\cite{18Zhang}. Due to the limited transmit power, the objective function of (\ref{OptC}) will be a monotonically non-decreasing sequence with an upper bound, which converges to a stationary solution that is at least locally optimal.}
\begin{algorithm}[t]
	\small{\caption{The Proposed Iterative Algorithm.}
		\label{algorithm2}
		{\bf{Initialize}} $\{a_{lk}^{[n]}, \xi_{lk}^{[n]},d_{lk}^{[n]},\tau_{lk}^{[n]},\beta_{lk}^{[n]},\upsilon_{lk}^{[n]},\lambda_{lm}^{[n]},\omega_{lm}^{[n]}\}$, $n=0$, the maximum iteration $I_{\rm{max}}$.\\
		\Repeat{$n=T_{\rm{max}}$ {\rm{or} Convergence (threshold=$10^{-3}$)}}{
			Update $n\leftarrow n+1$.\\
			Solve problem (\ref{OptC}) and obtain its optimal solution $\{{\bf{V}}_l^{[n]},{\bf{W}}_{l,k}^{[n]},\beta_{lk}^{[n]},a_{lk}^{[n]}, \xi_{lk}^{[n]},\xi_{lk}^{[n]},d_{lk}^{[n]},\tau_{lk}^{[n]},\beta_{lk}^{[n]},\upsilon_{lk}^{[n]},\lambda_{lm}^{[n]},\omega_{lm}^{[n]}\}$.}}
\end{algorithm} 

Now, we  analyze the complexity of the proposed algorithm. Given an iterative accuracy $\varsigma$, the number of iterations is on the order $\sqrt{\Delta}\ln(1/\varsigma)$, where $\Delta=14KL\!+\!4ML\!+MKL+\!NL\!+\!L\!+\!1$ denotes the barrier parameter related to the constraints~\cite{43Com}. In addition, (\ref{OptC}) includes $4KL+3ML+L+1$ liner constraints, $3KL$  two-dimensional LMI constraints, $L$ $N$-dimensional  LMI constraint, $KL$  $M$-dimensional  LMI  constraints and $2KL+ML$ second order cone constraints. Therefore, the total complexity of solving~(\ref{OptC}) is given by
%\begin{eqnarray}
${\mathcal{O}}\left(z\sqrt{\Delta}\ln(1/\varsigma)(z_1+z_2z+z_3+z^2)\right)$,
%\end{eqnarray} 
where $z={\mathcal{O}}(N^2L+KLM^2)$ and $N^2L+KLM^2$ denotes the number of decision variables,  $z_1=28KL+3ML+N^3L+M^3KL+L+1$, $z_2=15KL+3ML+N^2L+M^2KL+L+1$, and $z_3=KL((M+2)^2+(N+2)^2)+4KL$.

\section{Numerical Results}
In this section, numerical results are provided to evaluate the performance of our proposed algorithm. We assume that there are $L=2$ BS clusters, and each cluster includes $M=3$ BSs and $K=2$ users. The CP is equipped with $N=8$ antennas. {We assume that all users and BSs are uniformly distributed within a circular cell with 40 m radius. The distance between the CP and the BS cluster center is 300 m.} The mmWave bandwidth and microwave bandwidth are assumed $B_{\rm{mm}}=40$ MHz and $B_{\rm{mc}}=20$ MHz, respectively. The path loss is modeled as $69.7+24\log_{10}(d)$ dB at mmWave frequency and $38+30\log_{10}(d)$ dB at microwave frequency, where $d$ in meter is the distance~\cite{19Zhang}. The noise variance is set as -174 dBm/Hz, and the noise power caused by the ID at the users is -100 dBm. For simplicity, we set the minimum harvested energy for each user to be the same, and denote it by $E^{\rm{min}}$. Meanwhile, the maximum transmit power for each BS is also set the same, and denoted by $P_{\rm{AC}}^{\rm{max}}$. The energy conversion efficiency is set to $\eta=0.8$.

%\begin{figure}[t]
%	\begin{center}
%		\includegraphics[width=9cm,height=7cm]{Iteration.pdf}
%		\caption{Sum rate versus iteration.}
%		\label{Iteration}
%	\end{center}
%\end{figure}
%\begin{figure}[t]
%	\begin{center}
%		\includegraphics[width=9cm,height=7cm]{CPPower.pdf}
%		\caption{Sum rate versus maximum transmit power of the CP.}
%		\label{CPPower}
%	\end{center}
%\end{figure}
%\begin{figure}[t]
%	\begin{center}
%		\includegraphics[width=9cm,height=7cm]{BSPower.pdf}
%		\caption{Sum rate versus maximum transmit power of each BS.}
%		\label{BSPower}
%	\end{center}
%\end{figure}

Fig.~\ref{Iteration} shows the convergence performance of our proposed algorithm, where $E_{lk}^{\rm{min}}=1$ mW and $P_{\rm{AC}}^{\rm{max}}=30$ dBm.  One can observe that  the sum rate  converges after 15 iterations for $P_{\rm{CP}}^{\rm{max}}=40$ dBm, and  about 20 iterations for $P_{\rm{CP}}^{\rm{max}}=43$ dBm. Meanwhile, as expected, the sum rate is high when the CP's allowable transmit power is higher. This is because a higher CP's transmit power can provide a larger fronthaul rate. 

Fig.~\ref{CPPower} shows the sum rate versus maximum transmit power of the CP under different $E_{lk}^{\rm{min}}$ and $P_{\rm{AC}}^{\rm{max}}$. Under all considered conditions, the sum rate first increases with $P_{\rm{CP}}^{\rm{max}}$, and then saturates. In addition, although improving $P_{\rm{CP}}^{\rm{max}}$ can increase the fronthaul rate, this does not necessarily lead to the growth of the sum rate, since the latter is also affected by the transmit power of the BSs. For example, the sum rate has reached the maximum when $P_{\rm{CP}}^{\rm{max}}=43$ dBm and  $P_{\rm{AC}}^{\rm{max}}=30$ dBm, and it keeps a constant even for a higher $P_{\rm{CP}}^{\rm{max}}$. Furthermore, we can find when $P_{\rm{CP}}^{\rm{max}}$ is relatively low, the sum rate is the same for all considered conditions. This is because that the fronthaul rate determines the sum rate for a low  $P_{\rm{CP}}^{\rm{max}}$, while the BS's transmit power can simultaneously satisfy the requirement of harvested energy and maximum fronthaul rate. However, as   $P_{\rm{CP}}^{\rm{max}}$ increases, the access link rate determines the sum rate due to the limited transmit power of each BS. As a result, the sum rate is lower for a large $E_{lk}^{\rm{min}}$.

We plot how the sum rate varies  with $P_{\rm{AC}}^{\rm{max}}$ in~Fig.~\ref{BSPower}, where we set different $E_{lk}^{\rm{min}}$ and $P_{\rm{CP}}^{\rm{max}}$ for comparison. Similar to~Fig.~\ref{CPPower}, the sum rate  first increases and then remains stable when $P_{\rm{BS}}^{\rm{max}}$ increases. In fact,  when  $P_{\rm{BS}}^{\rm{max}}$ is low, the BSs first need to satisfy the requirement of each user's harvested energy  and then the remaining power can be used to transform data. Nonetheless, when  $P_{\rm{BS}}^{\rm{max}}$ is higher, the BSs have enough power to support the requirement of each user's harvested energy and maximum fronthaul rate provided by the CP. The  reasons are the same with that Fig.~\ref{CPPower}.
\section{Conclusion}
In this paper, we have investigated the joint beamforming and power splitting design problem in a BS cluster-based SWIPT C-RAN with multicast fronthaul. We have proposed a joint optimization algorithm of the multicast fronthaul beamforming, cooperative access beamforming and power split ratios to maximize the sum rate of the system. Meanwhile, we have analyzed the solution profile. Simulation results have verified the effectiveness of our proposed algorithm, and shown the effect of the CP's and BSs' transmit power on the sum rate. 
\appendix
\section{The Rank-One Reconstruction For ${\bf{V}}_0^\ast$ }\label{appendixA}
{Let ${{\bf{V}}_l^\ast}$ and ${{\bf{W}}_{lk}^\ast}$ denote the obtained solutions of problem (\ref{OptA}) via our proposed algorithm. If ${\rm{rank}}({{\bf{V}}_l^\ast})=1$ and ${\rm{rank}}({{\bf{W}}_{lk}^\ast})=1$, they can be respectively expressed as ${{\bf{V}}_l^\ast}=\theta_l{{\bf{v}}}_l^\ast({{\bf{v}}}_l^\ast)^H$ and ${{\bf{W}}_{lk}^\ast}=\theta_{lk}{{\bf{w}}}_{lk}^\ast({{\bf{w}}}_{lk}^\ast)^H$ by using  eigenvalue decomposition (EVD) method, and the optimal beamforming can be directly obtained as ${{\bf{v}}_l^{\rm{o}}}=\sqrt{\theta_l}{{\bf{v}}}_l^\ast$  and ${{\bf{w}}_{lk}^{\rm{o}}}=\sqrt{\theta_{lk}}{{\bf{w}}}_{lk}^\ast$. Otherwise, we adopt the randomization technique method (see, i.e.,~\cite{18Rank1}) to obtain the rank-one ${{\bf{V}}_l^\ast}$ and ${{\bf{W}}_{lk}^\ast}$. Specifically, applying the EVD technique, we decompose ${\bf{V}}_l^\ast={{\bf{X}}}_l{{\bf{D}}}_l({{\bf{X}}}_l)^H$ and ${\bf{W}}_{lk}^\ast={{\bf{Y}}}_{lk}{{\bf{\Omega}}}_{lk}({{\bf{Y}}}_{lk})^H$.
Then, by introducing the random vector ${{\bf{s}}}^{[n]}\sim {\mathcal{CN}}({\bf{0}},{\bf{I}})$ and ${{\bf{u}}}^{[n]}\sim {\mathcal{CN}}({\bf{0}},{\bf{I}})$, we form the $n$th candidate beamforming vector as ${{\bf{v}}}_l^{[n]}={{\bf{X}}}_l{{\bf{D}}}_l^{{1}/{2}}{{\bf{s}}}^{[n]}$ and ${{\bf{w}}}_{lk}^{[n]}={{\bf{Y}}}_{lk}{{\bf{\Omega}}_{lk}}^{{1}/{2}}{{\bf{u}}}^{[n]}$, respectively. As a result, we have $\mathbb{E}\{{{\bf{v}}}_l^{[n]}({{\bf{v}}}_l^{[n]})^H\}={\bf{V}}_l^\ast$ and $\mathbb{E}\{{{\bf{w}}}_{lk}^{[n]}({{\bf{w}}}_{lk}^{[n]})^H\}={\bf{V}}_{lk}^\ast$. Finally, we substitute the $n$th candidate beamforming vectors ${{\bf{v}}}_l^{[n]}$ and ${{\bf{w}}}_l^{[n]}$ into problem (\ref{OptA}) and reformulate the following optimization problem:
\begin{subequations}\label{OptD}
	\begin{align}
	\;\;\;\;\;\;\;\;\underset{\{{{c}}_l^{[n]},{{t}}_{lk}^{[n]},\beta_{lk}^{[n]}\}}{\rm{max}}\;&\sum\nolimits_{l=1}^{L}\sum\nolimits_{K=1}^{K}{R}_{lk}^{{\rm{AC}}[n]}\label{OptD0}\\
	{\rm{s.t.}}\;\;&{E}_{lk}^{[n]}\geq E_{lk}^{\min}, \forall l,k,\label{OptD1}\\
	&\sum\nolimits_{l=1}^{L}c_l^{[n]}||{\bf{v}}_l^{[n]}||^2\leq P_{\rm{CP}}^{\max}, \forall l,\label{OptD2}\\
	&\sum\nolimits_{k=1}^{K}t_{lk}^{[n]}|{\bf{w}}_{l,k}^{[n]}(m)|^2\leq P_{l,m}^{\max}, \forall l,m,\label{OptD3}\\
	&\sum\nolimits_{k=1}^{K}{R}^{{\rm{AC}}{[n]}}_{lk}\leq {R}_{l}^{{\rm{FH}}{[n]}}, \forall l,\label{OptD4}
	\end{align}
\end{subequations} 
where ${{c}}_l^{[n]}$ and ${{t}}_{lk}^{[n]}$ are coefficients,
${R}_{lk}^{{\rm{AC}}[n]}=B_{\rm{mm}}\log\left(1+\gamma_{lk}^{[n]}\right)$, $\gamma_{lk}^{[n]}=\frac{{{t}}_{lk}^{[n]}|{\bf{g}}_{llk}{{\bf{w}}}_{lk}^{[n]}|^2}{\sum_{i\neq k}^{K}{{t}}_{li}^{[n]}|{\bf{g}}_{llk}{{\bf{w}}}_{li}^{[n]}|^2+\sum_{j\neq l}^{L}\sum_{i=1}^{K}{{t}}_{ji}^{[n]}|{\bf{g}}_{jlk}{{\bf{w}}}_{ji}^{[n]}|^2+B_{\rm{mm}}\delta^2+\delta^2_u/\beta_{lk}^{[n]}}$, 
${R}_{l}^{{\rm{FH}}{[n]}}=\underset{m\in{\mathcal{M}}}{\min}\;\left\{R_{lm}^{{\rm{FH}}[n]}\right\}$, $E_{lk}=\eta(1-\beta_{lk}^{[n]})\left(\sum_{j=1}^{L}\sum_{i=1}^{K}{{t}}_{ji}^{[n]}|{\bf{g}}_{jlk}{\bf{w}}_{ji}^{[n]}|^2+B_{\rm{mm}}\delta^2\right)$
and $R_{lm}^{{\rm{FH}}[n]}=B_{\rm{mc}}\log\left(1+\frac{{{c}}_l^{[n]}|{\bf{h}}_{lm}{{\bf{v}}}_l^{[n]}|^2}{\sum_{j\neq l}^{L}{{c}}_j^{[n]}|{\bf{h}}_{lm}{{\bf{v}}}_j^{[n]}|^2+B_{\rm{mc}}\delta^2}\right)$.
It can be observed that problem (\ref{OptD}) can be solved using the iterative method proposed in Section III. Finally, we execute problem (24) repetitively for multiple different candidate vectors and select the optimal $c_l^{[n]\ast}$ and $t_{lk}^{[n]\ast}$ that owns the maximum sum rate, i.e.,~${\bf{v}}_l^{\rm{o}}=\sqrt{c_l^{[n]\ast}}{\bf{v}}_k^{{[n]}}$ and  ${\bf{w}}_{lk}^{\rm{o}}=\sqrt{t_{lk}^{[n]\ast}}{\bf{w}}_{lk}^{{[n]}}$.}

\end{document}